\documentclass[prd,preprint,nofootinbib]{revtex4}

\usepackage{graphicx, cancel}
\usepackage{epsf}
\usepackage{amsmath}
\usepackage{graphics}
\usepackage{amsfonts}
\usepackage{amssymb}
\usepackage{latexsym}
\usepackage{color}
\input{colordvi.tex}

\begin{document}
\renewcommand{\thefootnote}{\#\arabic{footnote}}
\newcommand{\rem}[1]{{\bf [#1]}}

\newcommand{\gsim}{ \mathop{}_ 
{\textstyle \sim}^{\textstyle >} }
\newcommand{\lsim}{ \mathop{}_ 
{\textstyle \sim}^{\textstyle <} }
\newcommand{\vev}[1]{ \left\langle {#1}  
\right\rangle }

\newcommand{\bear}{\begin{array}}  \newcommand 
{\eear}{\end{array}}
\newcommand{\bea}{\begin{eqnarray}}   
\newcommand{\eea}{\end{eqnarray}}
\newcommand{\beq}{\begin{equation}}   
\newcommand{\eeq}{\end{equation}}
\newcommand{\bef}{\begin{figure}}  \newcommand 
{\eef}{\end{figure}}
\newcommand{\bec}{\begin{center}}  \newcommand 
{\eec}{\end{center}}
\newcommand{\non}{\nonumber}  \newcommand 
{\eqn}[1]{\beq {#1}\eeq}
\newcommand{\la}{\left\langle}  
\newcommand{\ra}{\right\rangle}
\newcommand{\ds}{\displaystyle}

\def\SEC#1{Sec.~\ref{#1}}
\def\FIG#1{Fig.~\ref{#1}}
\def\EQ#1{Eq.~(\ref{#1})}
\def\EQS#1{Eqs.~(\ref{#1})}
\def\lrf#1#2{ \left(\frac{#1}{#2}\right)}
\def\lrfp#1#2#3{ \left(\frac{#1}{#2} 
\right)^{#3}}
\def\GEV#1{10^{#1}{\rm\,GeV}}
\def\MEV#1{10^{#1}{\rm\,MeV}}
\def\KEV#1{10^{#1}{\rm\,keV}}

\def\lrf#1#2{ \left(\frac{#1}{#2}\right)}
\def\lrfp#1#2#3{ \left(\frac{#1}{#2} 
\right)^{#3}}

\def\A{\mathcal{A}}
\renewcommand{\thefootnote}{\alph{footnote}}

\renewcommand{\thefootnote}{\fnsymbol{footnote}}
\preprint{IPMU 08-0022}
\title{A possible anthropic solution to the Strong CP problem}
\renewcommand{\thefootnote}{\alph{footnote}}

\author{Fuminobu Takahashi}

\affiliation{
 Institute for the Physics and Mathematics of the Universe,
   University of Tokyo, Chiba 277-8568, Japan
   }

\begin{abstract}
\noindent
We point out that the long-standing strong CP problem may be resolved
by an anthropic argument.  The key ideas are: 
(i) to allow explicit breaking(s) of the Peccei-Quinn symmetry which reduces 
the strong CP problem to the cosmological constant problem, 
and (ii) to conjecture that the probability distribution of the vacuum energy 
has a mild pressure towards higher values.  The cosmological problems of the 
(s)axion with a large Peccei-Quinn scale are absent in our mechanism,
since the axion acquires a large mass from the explicit breaking. 
\end{abstract}
\maketitle

%%%%%%%%%%%%%%%%%%%%
\section{Introduction}
\label{sec:1}
%%%%%%%%%%%%%%%%%%%%
One of the profound problems of the standard model (SM) is the
strong CP problem. In the quantum chromodynamics (QCD),
there is no a priori reason to forbid the following CP-violating
operator,
\begin{eqnarray}
    \mathcal{L} \;=\; \frac{g_s^2 \,\theta}{64\pi^2} 
    \epsilon_{\mu\nu\rho\sigma} 
    G^{(a)\mu\nu} G^{(a)\rho\sigma},
\label{eq:QCDtheta}  
\end{eqnarray}
where $G^{(a)}_{\mu\nu}$ is the field strength of the $SU(3)_c$ gauge
fields, and $g_s$ is the $SU(3)_c$ gauge coupling.  This operator
contributes to the electric dipole moment of the neutron, and the
experimental measurements have severely limited the parameter $\theta$
as $|\theta| < 10^{-(9-10)} \equiv \theta^{(\rm
exp)}$~\cite{Baker:2006ts}. Such a tight constraint on $\theta$ is regarded as a fine-tuning; this is 
the strong CP problem.

The Peccei-Quinn (PQ) mechanism provides a natural
solution to the strong CP problem~\cite{Peccei:1977hh}. In the
mechanism, one introduces an axion~\cite{Peccei:1977hh,
Weinberg:1977ma, Wilczek:1977pj}, which is charged under
the PQ symmetry. Under the PQ transformation, the axion field  $a$ gets
shifted as $a \rightarrow a + f_a\, \epsilon$, where $f_a$ denotes the
axion decay constant (or the PQ scale), and $\epsilon$ is the
transformation parameter.  
In what follows we normalize
the axion $a$ by $f_a$ so that $a$ is dimensionless.
The axion is assumed to be coupled
to the QCD anomaly,
\begin{eqnarray}
    \mathcal{L} \;=\; \frac{g_s^2}{64\pi^2}\,a\,
    \epsilon_{\mu\nu\rho\sigma} 
    G^{(a)\mu\nu} G^{(a)\rho\sigma}.
    \label{eq:axion-gluon}
\end{eqnarray}
After the QCD phase transition, the axion gets stabilized due to the
QCD instanton effect, satisfying $a + \theta = 0$. Thus the strong CP
problem is solved dynamically.

Since the PQ mechanism was proposed, a lot of efforts have been made to
implement the mechanism. The models
proposed so far can be divided broadly into two categories. One adopts
a field theoretic approach using a $U(1)_{\rm PQ}$
symmetry. In the DFSZ~\cite{Dine:1981rt,Zhitnitsky:1980tq} and
KSVZ (or hadronic)~\cite{Kim:1979if,Shifman:1979if} axion models,  
a global $U(1)_{PQ}$ symmetry is introduced, which is
spontaneously broken by a vacuum expectation value (VEV) of a scalar
field.  The associated Nambu-Goldstone  boson becomes an axion.  
Those models fall in this category.  The other identifies one of the axion-like fields 
in the string theory to be the QCD axion. We focus on the latter category throughout this letter.

The string theory is currently the most promising candidate
for a unified theory of all forces including gravity~\cite{string}.
Moreover, it contains many axion-like fields associated with the
Green-Schwarz mechanism~\cite{Green:1984sg}. Therefore, it is natural
to seek for the QCD axion in the string set-up. However, it turns out that 
there are severe cosmological problems associated with the axion.

The PQ scale $f_a$ is constrained as $10^9~{\rm GeV} \lesssim f_a \lesssim
10^{12}~{\rm GeV}$~\cite{Raffelt:1996wa,Kolb:1990vq,Preskill:1982cy}
from astrophysical and cosmological considerations. The upper bound
comes from the requirement that the axion density should not exceed
the observed amount of dark matter (DM), based on an assumption that the
initial displacement of the axion from the nearest minimum is ${\cal
O}(1)$.  However, the PQ scale is expected to be as large 
as ${\cal O}(10^{16})$\,GeV in the string theory.
If $f_a$ is as large as $10^{16}\,$GeV, the axion abundance would
exceed the observed DM abundance by many orders of magnitudes.
Although we may hope that the axion model with smaller $f_a$ can be constructed, 
currently it seems hard to make the value of $f_a$ much smaller than 
$10^{16}{\rm\,GeV}$~\cite{Svrcek:2006yi}.  
There are several solutions proposed so far; (i) to dilute the axion
abundance by the late-time entropy
production~\cite{Kawasaki:1995vt}; (ii) to set the initial position of
the axion very close to the CP conserving minimum. However both are
not completely satisfactory.

The first solution (i) is most easily realized by introducing the late-time
decaying particle~\cite{Lyth:1995ka} or unstable topological
defects~\cite{Kawasaki:2004rx}, which produce enormous amount of the
entropy at the decay. However, since the pre-existing baryon asymmetry is also
diluted, we have to rely on a very efficient baryogenesis scenario such
as the Affleck-Dine mechanism~\cite{AD,
DRT,Stewart:1996ai,Kasuya:2001tp,Jeong:2004hy,Kawasaki:2006py}.  We do
not argue that it is impossible to have consistent cosmology in this
case, but  the cosmology required by this solution is far from the simplest 
one, making us feel that it is slightly contrived.

In the second solution (ii), we need to fine-tune the initial position
of the axion. Since the axion likely  takes a randomly
chosen value due to quantum fluctuations during inflation,
we need to indeed fine-tune the initial position by hand.
One may hope that the initial position of the axion might be 
selected in such a way that the
axion abundance does not exceed the DM
abundance~\cite{Linde:1987bx}, based on the anthropic principle.
When applied to the cosmological constant, the anthropic principle
was successful as shown in~\cite{Weinberg:1987dv}.  
However, the recent analysis showed that the constraint
on the DM abundance, therefore on the initial position of the axion, 
is too loose based on the simple anthropic argument~\cite{Hellerman:2005yi}.  
On the other hand, the authors of Ref.~\cite{Tegmark:2005dy} performed 
much more detailed studies by taking account of e.g. the comet impact rate
in a universe with a larger amount of dark matter. Their results showed that
the anthopically favored value of the dark matter abundance is very close
to the observed one. While we agree that the comet impact rate can have an
important effect on the existence of life, it is not easy to estimate its effect
precisely due to our limited knowledge.

The bosonic supersymmetric (SUSY) partner of the axion, saxion, also leads to a severe cosmological
problem~\cite{Banks:1996ea,Banks:2002sd,Kawasaki:2007mk}, which is
similar to the notorious cosmological moduli
problem~\cite{ModuliProblem,MGP,MGP2,Endo:2006tf}.  One may be able to
solve the problem  in a similar fashion described
above, but the resultant cosmology again does not seem natural.
Note also that the anthropic argument on the (s)axion abundance cannot solve the problem,
unless the saxion is stable in cosmological time. 

While a starting point is well grounded theoretically, i.e., the
axion elegantly solves the strong CP problem and the string theory
seems to be the plausible candidate to implement the PQ mechanism, we
are nevertheless led to either apparently contrived cosmology or the
fine-tuning.
Those tantalizing situation can be viewed as a hint that we might have
made a wrong assumption from the very beginning. That is to say, the
dynamical solution to the strong CP problem may not be the correct
answer, if the axion is to be embedded in the string theory.  

 In this letter, we give up the ordinary PQ mechanism, and instead, we
consider what happens if the PQ symmetry is explicitly broken other
than the QCD instantons. The beauty of the PQ mechanism has prevented
most people to pursue this possibility seriously.  We
find that the CP conserving minimum can be anthropically selected, if
the probability distribution of the vacuum energy excluding the contribution
from the axion sector has a pressure toward higher values. Whether the
probability distribution possesses such a property or not 
is tied to the cosmological measure problem, and we do not have a definite answer 
at the moment. We will, however, give several possibilities that such a feature may appear.

 It is quite interesting to note that the axion can acquire a large mass due
to the explicit breaking, and it may be absent in the low-energy
particle spectrum~\footnote{
We use the terminology, ``axion", although it is not the ordinary massless QCD axion in the
PQ mechanism.
}.  This striking feature has rich implications for
cosmology.  All the cosmological problems associated with the (s)axion
are solved, if the (s)axion mass is large enough. The axion may come
to dominate the energy density of the universe after inflation, and
reheat the universe by its decay. It is even possible to make the
cosmological abundance of the axion negligible, if the explicit
breaking is large enough during inflation.

To summarize, with our conjecture on the probability distribution of
the vacuum energy, we arrive at the
followings.
\begin{enumerate}
\item  The strong CP problem is resolved by the anthropic reasoning.
\item The cosmological problems
of the (s)axion with large $f_a$ can be solved.
\item  Interesting cosmological
scenarios emerge: the axion may dominate and reheat the universe; the
axion may generate the cosmological density perturbations.
\end{enumerate}
In the following sections, we will detail each point.

%%%%%%%%%%%%%%%%%%%%%%%%%%%%%%%%%%%%%%%%
\section{The anthropic solution to the strong CP problem}
\label{sec:2}
%%%%%%%%%%%%%%%%%%%%%%%%%%%%%%%%%%%%%%%%
Now let us explain how it works. The shift symmetry of the axion is
violated by the QCD instantons. After the QCD phase transition,
the instantons generate the effective potential of the axion,
\beq
V_{\rm QCD}(a)\;=\;\Lambda_{\rm QCD}^4 \left(1-\cos{a}\right),
\label{eq:pq}
\eeq
where the axion field $a$ is dimensionless, and we have chosen the CP
conserving minimum to be at $a=0$ for simplicity. We drop numerical
coefficients of order unity here and in what follows, since they are
irrelevant for our discussion.  If there are no other contributions to
the axion potential, the axion will settle down to $a = 0$ after the
QCD phase transition, and the strong CP problem is dynamically solved.

\begin{figure}[t]
\begin{center}
\includegraphics[width=10cm]{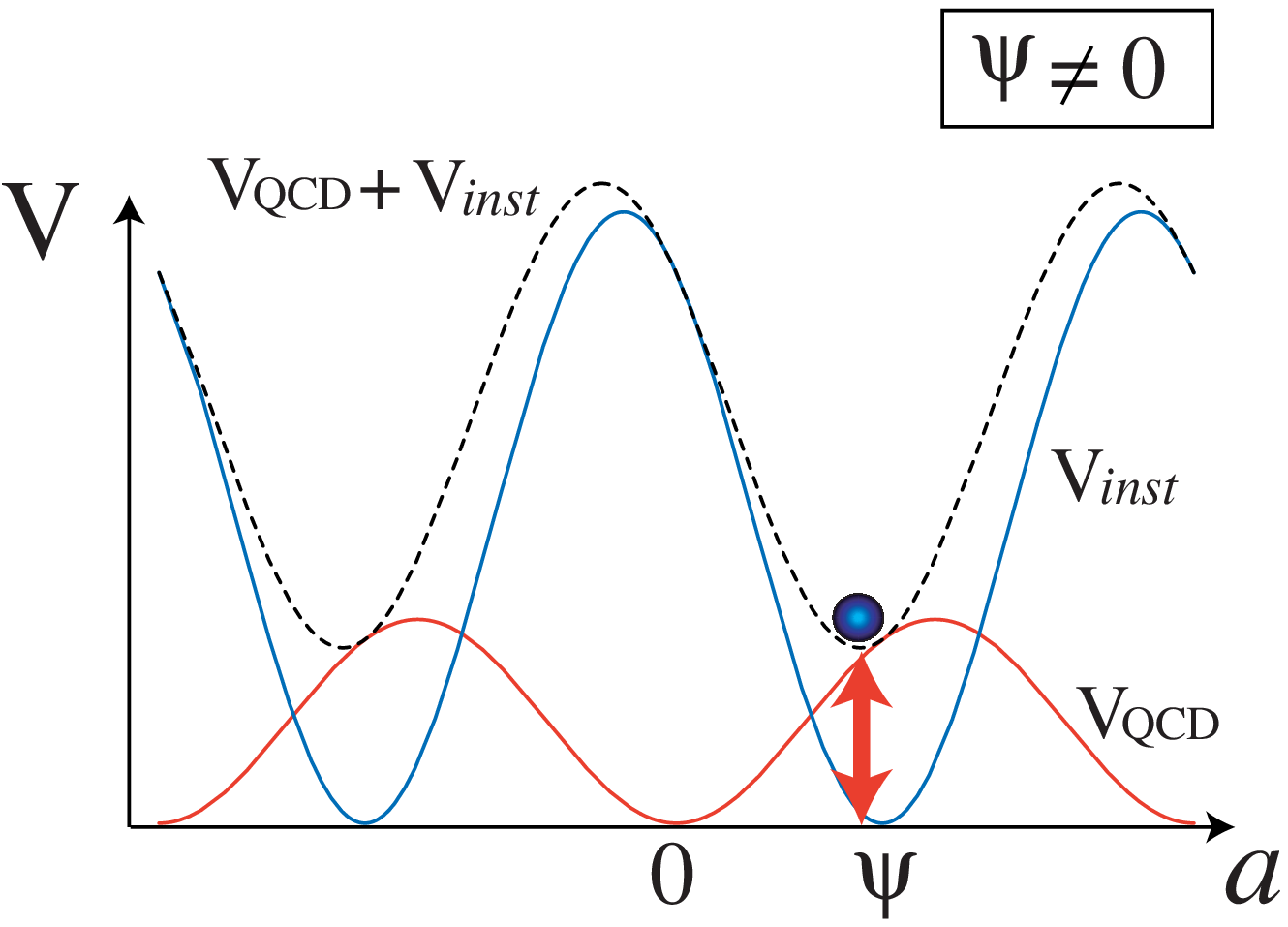}
\includegraphics[width=10cm]{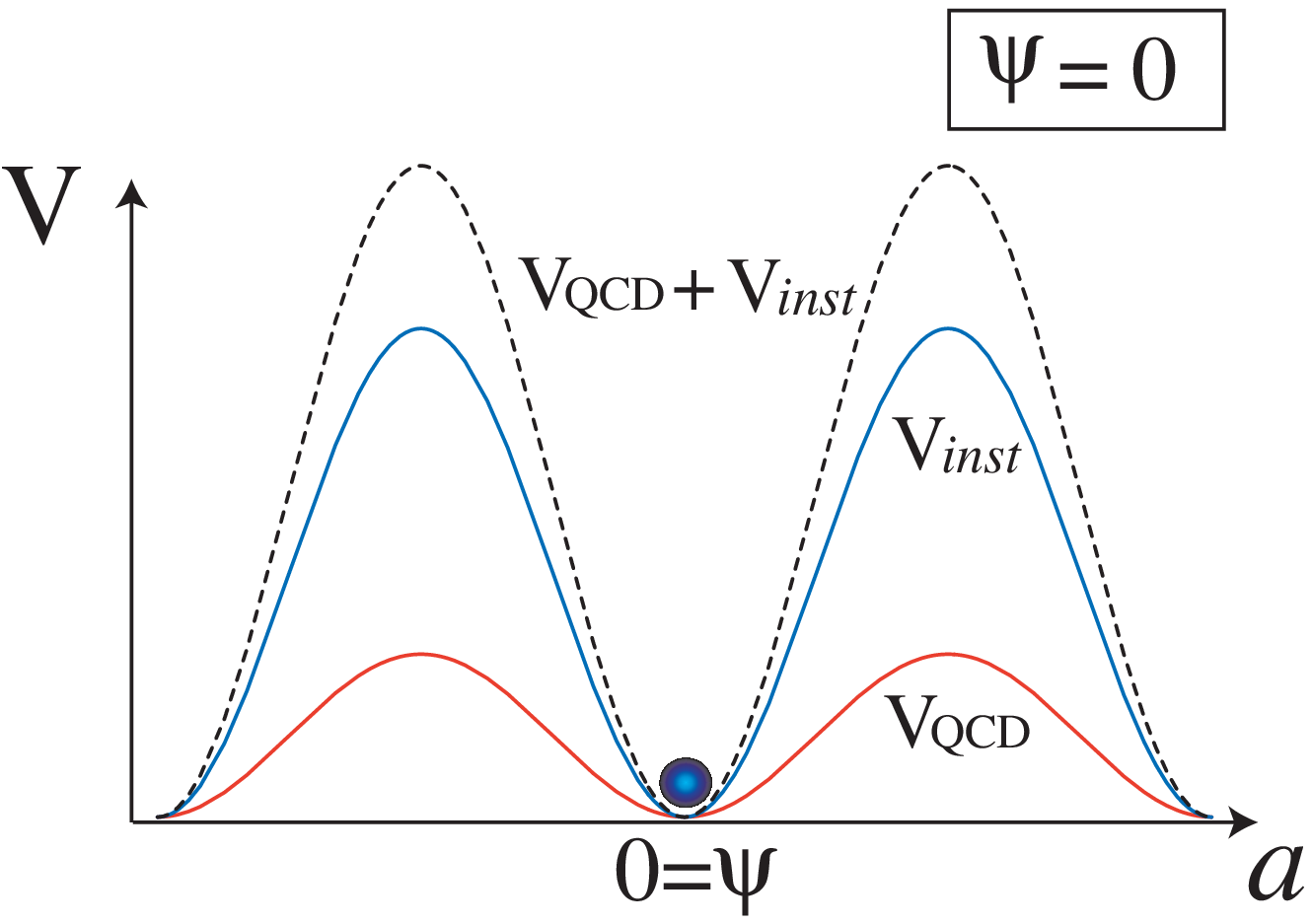}
\caption{The axion potentials, $V_{\rm QCD}$,
$V_{inst}$, and $V_{\rm QCD} +V_{inst}$,
for $\psi \ne 0$ (top) and $\psi = 0$ (bottom). 
The circle represents the minimum
of the axion potential, and the arrow shows non-zero
cosmological constant at the minimum.  }
\label{axion}
\end{center}
\end{figure}

Let us  introduce another explicit breaking of the shift symmetry,
which generates the following potential,
\beq
V_{inst}(a)\;=\; \Lambda_{inst}^4 \left(1-\cos{\left(a-\psi\right)}\right)
\label{eq:modified-pq}
\eeq
where $\psi$ denotes the minimum of the explicit breaking term.
Indeed, there is such breaking of the shift symmetry due to some sort
of the instantons in the string theory~\cite{Svrcek:2006yi}.  The
precise form of the explicit breaking is not important here.  
How the explicit breaking is generated and how large it is will be 
discussed later. For the
moment we assume that the potential $V_{inst}$ is the only
source for the explicit breaking of the shift symmetry, other than the
QCD instanton.   The total axion potential after the QCD phase transition 
is given by $V(a) = V_{\rm QCD}(a)+V_{inst}(a)$.  See Fig.~\ref{axion}.  We assume that the
breaking term is much larger than the term arising from the QCD
instantons, i.e.,
\beq
\Lambda_{inst} \;\gg\; \Lambda_{\rm QCD}.
\eeq
Then the minimum of the axion potential $V(a)$ is essentially
determined by that of $V_{inst}(a)$. That is,
$V(a)$ takes the minimal value 
at $a \approx \psi$.
Generically we expect $\psi = {\cal
O}(1)$, because there is no a priori reason for the explicit breaking
term to have its minimum just at the CP conserving one. Therefore, we
have intolerably large CP phase in the presence of the large explicit
breaking of the PQ symmetry, as expected.  That is why we usually
assume that such explicit breaking is somehow suppressed for the PQ
mechanism to work.

We now assume that $\psi$ is an environmental variable, which takes 
different values in different regions in the universe that are
separated far apart from one another. One can imagine a situation that
there are an infinitely large number of expanding regions, in each of
which $\psi$ takes a different value. 

We note that the axion potential
at the  minimum becomes the smallest when $\psi$ equals to
$0$, i.e., when the minimum of $V_{inst}$ happens to coincide with
that of $V_{\rm QCD}$ (see the bottom panel in Fig.~\ref{axion}).
In this sense the CP conserving minimum is special. However,
since there are many other contributions to the total cosmological
constant, one cannot naively argue that the CP conserving minimum
with $\psi = 0$ should be selected simply because it minimizes the
contribution from the axion sector. 

To illustrate our idea, let us consider such a universe with $\psi = 0$
that the total cosmological constant including all the possible 
contributions is within the anthropic window, i.e.,  the
cosmological constant  is small enough to make the universe habitable.  
Below we will discuss a condition that such a universe becomes more 
likely than the others.  For the moment, let us consider 
what happens if we vary $\psi$ from $0$ in such a universe. 
Then, even  if $\psi$ were slightly different from $0$, the cosmological constant
 would be greatly enhanced as $\rho_{\rm cc} \approx \Lambda_{\rm
 QCD}^4\, |\psi|^2$ for $|\psi| \ll 1$, and it would be out of the
 anthropic window. Here $\rho_{\rm cc}$ denotes the energy density of
 the cosmological constant, including all the contributions.  Thus,
life cannot arise in such universe with $\psi = {\cal O}(1)$, and
 the universe with almost vanishing $\psi$ will be selected by the
 anthropic principle. It is worth mentioning here that the explicit
 breaking of the PQ symmetry has reduced the strong CP problem to
the cosmological constant problem.

The remaining issue is why the total cosmological constant should be almost
zero (more precisely, within the anthropic window) in the universe
with $|\psi| \approx 0$.  In other words, among those universes with the total cosmological
constant satisfying the anthropic bound, is there any reason to favor
smaller values of $|\psi|$? We here adopt a conjecture that there are
infinitely large number of meta-stable vacua, in each of which the cosmological
constant takes a variety of values, i.e., the so-called string
landscape~\cite{Lerche:1987sg,Bousso:2000xa,Kachru:2003aw}.  To be
explicit, we express the energy density of the cosmological constant
as follows:
\beq
\rho_{\rm cc} \;=\;  \rho_{ L} + \rho_{\rm axion}(\psi)
\label{cc}
\eeq
with
\beq
\rho_{\rm axion}(\psi) \;\equiv\; \left.V(a)\right|_{a = \psi}.
\eeq
The first term in Eq.~(\ref{cc}), $ \rho_{ L} $, is supposed
to contain all the contributions such as the quantum corrections, the
electroweak symmetry breaking and the string landscape, except for
the axion potential, which is represented by the second term, $\rho_{\rm
axion}(\psi)$.  $ \rho_{ L} $ can take a variety of values in
the huge number of vacua, which enables us to live in such a universe
that the first and second terms (almost) cancel with each other,
giving $\rho_{\rm cc} \approx 0$. In the following we assume that the
main effects of varying $\psi$ is to change $V(a)$, i.e.,
$\rho_{\rm axion}(\psi)$.  More precisely, if we change $\psi$ with
all the other parameters being fixed, the change in $\rho_{\rm cc}$ is
assumed to be dominantly given by the change in $\rho_{\rm
axion}(\psi)$. 

One of the interesting features of the string landscape is that one
can in principle quantify the naturalness in terms  of probability 
by e.g. counting the number and/or weighing the volume of the vacua 
satisfying certain conditions of interest~\cite{Linde:1986fd,Linde:2006nw}.  
Let us define the probability distribution $P_{ L}(\rho_{ L})$
in such a way that the probability that a vacuum has
$\rho_{ L}$ in the range of $\rho_{ L} \sim \rho_{ L}
 + \Delta \rho_{ L}$ is 
given by $P_{ L}(\rho_{ L}) \Delta \rho_{ L}$. 
The probability is assumed to include
not only the a priori probability distribution, but also the other effects such as
the volume due to the eternal inflation and 
the statistical (or dynamical) properties of scanning the landscape.

We assume that the probability distribution of $\psi$ is flat for simplicity. 
Then the resultant probability distribution of $\rho_{\rm axion}$ is 
given by
\beq
P_{\rm axion}(\rho_{\rm axion}) \;=\; 
\left\{
\bear{cl}
\ds{\frac{1}{\pi \Lambda_{\rm QCD}^4} \left[2 \left(\frac{\rho_{\rm axion}}{\Lambda_{\rm QCD}^4}\right) -\left(\frac{\rho_{\rm axion}}{\Lambda_{\rm QCD}^4}\right) ^2 \right]^{-\frac{1}{2}}} &{\rm ~for~~} 0<\rho_{\rm axion} < 2 \Lambda_{\rm QCD}^4,\\
&\\
0&{\rm ~~otherwise}.
\eear
\right.
\eeq
Note that the maximum and minimum are not specially favored, although $P_{\rm axion}(\rho_{\rm axion})$ diverges at $\rho_{\rm axion} = 0$ 
and $2 \Lambda_{\rm QCD}^4$; the probability remains finite.

The total cosmological constant $\rho_{\rm cc}$ is given by the sum of
$ \rho_{ L} $ and $\rho_{\rm axion}$, as shown in Eq.~(\ref{cc}). Therefore we naively expect that
there are many ways to make the first and the second terms almost cancel with
each other so that the total cosmological constant $\rho_{\rm cc}$
is within the anthropic window, $0<\rho_{\rm cc} \lesssim \rho_{\rm cc}^{\rm (aw)}
= {\cal O}((1{\rm\,meV})^4)$.

First, let us consider a case that $P_{ L}(\rho_L)$ is independent of
$\rho_L$ over an interested range of $ \rho_{ L} $.  We call
this case as the flat distribution. Then,  a vacuum 
satisfying  $0<\rho_{\rm cc} \lesssim \rho_{\rm cc}^{\rm (aw)}$ does not  favor
any particular value of $\psi$.
Whatever value $\psi$ takes, there are  some fixed number of
vacua that makes $\rho_{cc}$ almost zero.  In this sense, the universe
with $\psi=0$ is as likely as that with e.g. $\psi = 1$.  Therefore, if
the probability distribution of $ \rho_{ L} $ is flat over an interested range
of $ \rho_{ L} $, one cannot solve the strong CP problem by
the anthropic reasoning.

\begin{figure}[h!]
\begin{center}
\includegraphics[width=8cm]{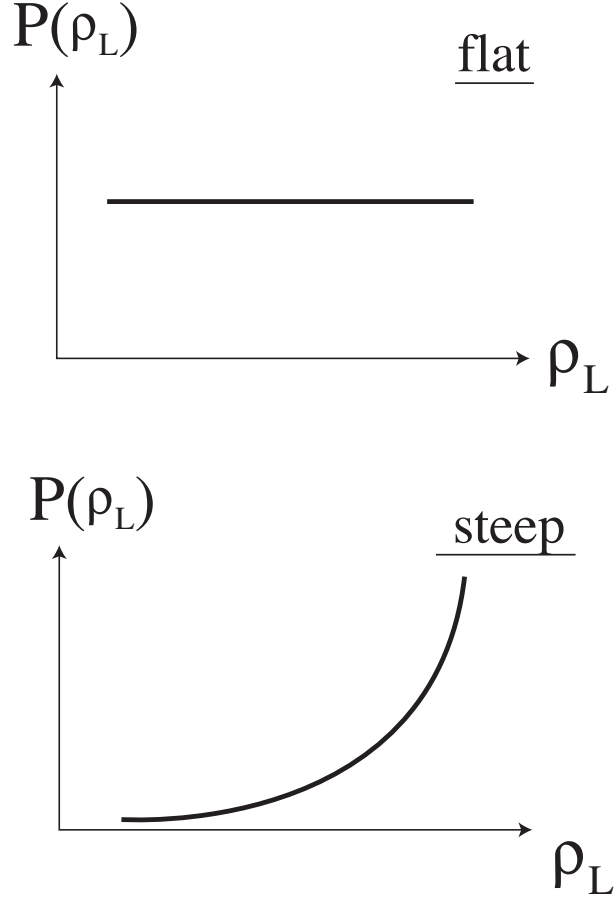}
\caption{The probability distributions of $ \rho_{ L} $. The flat (top)
and steep (bottom) distributions are shown.
The range of the distribution shown in this figure is supposed to be at least
$\sim \Lambda_{\rm QCD}^4$. We do not need to assume global behavior of $P_L$. 
 (see also footnote~\ref{footnote:c})
}
\label{dist}
\end{center}
\end{figure}

The situation greatly changes if we allow $P_L(\rho_L)$ to 
depend on $\rho_L$. Suppose that $P_L(\rho_L)$ grows  as 
$\rho_L$ increases. We call this case as the steep distribution.  For
instance, we can imagine an exponential form, $P_L(\rho_L) = P_0
\exp(\rho_L/\rho_0)$.  See Fig.~\ref{dist}.  Then, among those vacua
satisfying the anthropic bound, 
$0<\rho_{\rm cc} \lesssim \rho_{\rm cc}^{\rm (aw)}$, smaller values of
$|\psi|$ are favored, i.e., the universe with $\psi \approx 0$ is more likely
than that with e.g. $\psi = 1$.
 This is simply because the probability distribution $P_L(\rho_L)$
 is enhanced as $ \rho_{ L} $ increases, i.e., as $\rho_{\rm axion}$ decreases.
Note that the anthropic bound requires
\beq
0< \rho_L + \rho_{\rm axion} \;\lesssim\; \rho_{\rm cc}^{\rm (aw)}.
\label{anth}
\eeq
With this bound satisfied, making  $\rho_L$ larger is equivalent to 
making $\rho_{\rm axion}$ smaller. Due to the steep distribution,
larger $\rho_L$, or equivalently, smaller $\rho_{\rm axion}$ is favored.
 Thus, if the probability distribution
of $ \rho_{ L} $ is steep enough, the universe with
$\psi \approx 0$ is statistically favored among the universes
satisfying the anthropic constraint on
the cosmological constant.

Let us evaluate how steep  $P_L(\rho_L)$ should be to solve the
strong CP problem. To satisfy the current bound on $\theta$, $\psi$
must be constrained as $|\psi| < \theta^{(\rm exp)} = 10^{-(9-10)}$.
For $|\psi| \ll 1$, we have approximately
\beq
\rho_{\rm axion}(\psi) \;\simeq\; V_{\rm QCD}(\psi) 
\;\simeq \;\Lambda_{\rm QCD}^4 \frac{\psi^2}{2}.
\eeq
In order to statistically favor the universe with $|\psi| <
\theta^{(\rm exp)}$ over the universe with $|\psi| > \theta^{(\rm
exp)}$, the following condition must be met;
\beq
\int_{\rho_{\rm cc}^{\rm (aw)} - \frac{1}{2}\theta^{(\rm exp)2}
\Lambda_{\rm QCD}^4}^{\rho_{\rm cc}^{\rm (aw)}} P_L(\rho_L) \,d\rho_L
\;\gtrsim \;\int^{\rho_{\rm cc}^{\rm (aw)} - \frac{1}{2} \theta^{(\rm exp)2}
\Lambda_{\rm QCD}^4}_{\rho_{\rm cc}^{\rm (aw)}-\Lambda_{\rm QCD}^4} 
P_L(\rho_L) \,d\rho_L.
\label{condition}
\eeq
If one adopts an exponential form, $P_L(\rho_L) = P_0
\exp(\rho_L/\rho_0)$, the condition amounts to
\beq
\rho_0 \;\lesssim\; \theta^{(\rm exp)2} \Lambda_{\rm QCD}^4 \;\approx\; (1{\rm\,keV})^4.
\eeq
We would like to emphasize that such a steep distribution does not
have to persist over an entire range of $ \rho_{ L} $.  If the probability
distribution is locally steep at $ \rho_{ L}  \approx
\rho_{\rm cc}^{{\rm (aw)}} - \rho_{\rm axion}$ 
over a range of $\sim \Lambda_{\rm QCD}^4$, our
arguments above is valid~\footnote{
\label{footnote:c}
In the presence of multiple 
breaking terms with the strengths, $\Lambda_1^4 \ll 
 \cdots \ll\Lambda_{n-1}^4 \ll \Lambda_n^4$, with the relative differences of
 the minima being the environmental variables,
we need to assume that the steep distribution persists at least over
a range of $\Lambda_{n-1}^4$.
}.

Several remarks are as follows.  We have  assumed  that
the probability distribution of $\psi$ is almost flat. However, it is
not necessarily flat; it can depend on $\psi$ as long as the
dependence is mild enough that the argument above using the steepness
in  $P_L$ remains valid.
We can also imagine that $\theta$ as well may be an environmental
variable.  In this case, the above argument remains unchanged by
simply replacing $\psi$ with $\psi' \equiv \psi-\theta$. If there are multiple
scalars that couple to the QCD anomaly, we take our axion as the lightest
one, while the others are integrated out. Then the possible effects
of the heavier particles can be represented by varying $\theta$. 
One may
wonder if the axion is not introduced from the beginning but
$\theta$ is still regarded as an environmental parameter. One can
reach the same conclusion, since this essentially corresponds to the
case that the axion is integrated out.
 
 One may worry that the required steepness in the probability distribution of
 $ \rho_{ L} $ may contradict with the flat prior that is usually assumed when
 one applies the anthropic principle to the cosmological constant problem. 
  Both can be reconciled if the steepness is rather weak over the
 typical value of the cosmological constant within the anthropic
 window, but still strong enough to select the universe with $\psi
 \approx 0$.  In the case of the exponential form, this is satisfied
 if
\beq
\rho_{\rm cc}^{{\rm (aw)}} \;\ll\; \rho_0  \;\lesssim\; 
\theta^{(\rm exp)2} \Lambda_{\rm QCD}^4,
\eeq
or equivalently, $(1 {\rm\, meV})^4 \;\ll\; \rho_0  \;\lesssim\;  (1{\rm\,keV})^4$,
is met. 
%It is usually argued that the a priori probability distribution of the
%cosmological constant must be flat over the anthropic window, since
%the scale of $\sim 1\,$meV is much smaller than any fundamental physical 
%scales. In this sense, the QCD scale, which essentially determines the
%needed steepness, may be regarded as an important physical scale
%that determines the probability distribution in the landscape. 

What is the possible origin of the hierarchical distribution of  $ \rho_{ L} $?
Such a steep distribution is just a trade-off with the fine-tuning
of the initial position of the axion, until we find its origin.
Interestingly, however, there are several proposals 
 that the probability distribution of $ \rho_{ L} $ might
differ from the flat distribution~\cite{ArkaniHamed:2005yv,SchwartzPerlov:2006hi,SchwartzPerlov:2006hz,Olum:2007yk,Podolsky:2007vg,Podolsky:2008du}. 
In any case, this is closely related to the cosmological measure problem,
which is not settled yet~\footnote{We thank R. Bousso for discussion
and the healthy criticism of such steep distribution.}. 

Here we give one possible explanation for
the steep distribution. We assume that there are many meta-stable vacua with different values
of the cosmological constant. The universe trapped in a false vacuum with a large cosmological
constant will experience eternal inflation. After a long time, a bubble will be
created with its center at a vacuum with smaller cosmological constant.
One might expect that the most rapidly-inflating vacuum gives the dominant probability,
since it has a exponentially  large volume. 
Although it is not easy to define the gauge-invariant measure that rewards the volume,
we here assume that such a measure can be defined properly. 
Note that the QCD instanton effect is suppressed before the QCD phase transition.
Therefore, in an epoch much before the QCD phase transition, 
there is a difference in the energy density between the vacuum $``A"$ with $\psi = 0$ 
and the vacuum $``B"$ with $\psi= 10^{-9}$, as long as both satisfy the anthropic bound (\ref{anth}).
The energy difference, $\rho_A - \rho_B$, will be of ${\cal O}({\rm keV}^4)$.
We assume that all the other parameters other than $\psi$ are fixed.
The energy scale of ${\cal O}({\rm keV}^4)$ is small compared to the energy density of
the universe before BBN, which makes difficult to distinguish the two vacua $A$ and $B$
by the ordinary cosmological evolution. Suppose that there are first order phase transitions from
the long-lived meta-stable vacua, $A'$ and $B'$ into the vacua $A$ and $B$, respectively.
The energy densities of $A'$ and $B'$ are denoted by
$\rho_{A'}$ and $\rho_{B'}$.  We are considering such a situation
that the vacuum $A' (B')$ is  adjacent to the vacuum $A(B)$, while the vacua $A'$ and $B'$, therefore $A$ and $B$,  are far apart
from each other. We assume that a typical (or averaged) energy difference between the vacua $A$ and $A'$ is equal to
that between $B$ and $B'$, since ${\cal O}({\rm keV}^4)$ is so small compared to the fundamental scale.
It means that $\rho_{A'}$ tends to be slightly larger than $\rho_{B'}$ by ${\cal O}({\rm keV}^4)$.
The difference of the energy densities, $\rho_{A'} - \rho_A$ as well as $\rho_{B'}-\rho_B$,
takes a variety of values with probably large dispersion. So, one has to collect at least
$((M_P/{\rm keV})^4)^2 \sim 10^{195}$ vacua, in order to see such a tiny difference.
Of course, since ${\cal O}({\rm keV}^4)$ is much smaller than the fundamental scale, it does not affect
the cosmological expansion in most cases. However, if the meta-stable vacua $A'$ and $B'$
are very long-lived, say, if the typical number of the e-foldings during the inflation in the
vacua $A'$ and $B'$ is exponentially large, such a tiny difference in the energy density
may result in significant difference in the final volume. Thus, we may have a steep distribution $P_L$,
which changes over a scale of ${\cal O}({\rm keV}^4)$.
Note that, in the above explanation, the origin of the steepness is the huge number of the vacua
and the longevity of the meta-stable vacua $A'$ and $B'$.

So far we have assumed that the explicit breaking is much larger than
the QCD instanton effects. For our arguments to be valid, the axion
abundance should not contribute to the DM abundance. If it does, we
need to perform analysis along the line of
Ref.~\cite{Hellerman:2005yi}, and we will typically end up with the DM
abundance much larger than the observed value.  Therefore the explicit
breaking is assumed to be large enough that the axion does not
contribute to the DM abundance.  If there are several explicit
breaking terms with different strength, this restriction on the size
applies to the largest one. The anthropic argument can be similarly
applied to the smaller breaking terms. In particular it is no problem
to apply to the breaking terms smaller than the QCD instanton effects,
as long as the breaking is much larger than $\sim (1{\rm meV})^4$.

%%%%%%%%%%%%%%%%%%%%
\section{Cosmology}
\label{sec:3}
%%%%%%%%%%%%%%%%%%%%
In the ordinary PQ mechanism, the axion acquires its mass mainly from
the QCD instanton effects represented by (\ref{eq:pq}), and the mass
is given by
\begin{eqnarray}
    m_a 
    \;\sim\; m_\pi \frac{F_\pi}{f_a} 
    \;\simeq\; 1 \times 10^{-9}\,{\rm eV} 
    \lrfp{f_a}{10^{16}\,{\rm GeV}}{-1},
    \label{eq:axion-mass}
\end{eqnarray}
where the numerical coefficient weakly depends on the axion models.
Thus the axion is usually very light and stable, and that is why the
axion is one of the candidates for the DM.  In our scenario, however,
the axion acquires a large mass due to the explicit breaking of the PQ
symmetry. Assuming the breaking term given by (\ref{eq:modified-pq}),
the axion mass is
\beq
m_a \;\sim\; \frac{\Lambda_{inst}^2}{f_a}.
% 10^{8}\,{\rm GeV} 
% \lrfp{\Lambda_{inst}}{10^{12}\,{\rm GeV}}{2}   
% \lrfp{f_a}{10^{16}\,{\rm GeV}}{-1}.
 \eeq
So, the axion mass sensitively depends on  $\Lambda_{inst}$. 

How large is $\Lambda_{inst}$? In the string theory, there are several
sources for the explicit breaking of the shift symmetry: the
world-sheet instantons, brane instantons, gauge instantons from other
factors of the gauge group, and gravitational
instantons~\cite{Svrcek:2006yi}.  Since all of them are
non-perturbative effects, $\Lambda_{inst}$ can be exponentially
suppressed relative to the fundamental scale, $M$, which can be as
large as the reduced Planck scale, $M_P = 2.4 \times
10^{18}$\,GeV. That is, we estimate $\Lambda_{inst}^4 = M^4
\exp(-S_{inst})$, where $S_{inst}$ denotes the action of the
instanton. Or, in the presence of low-energy SUSY, it might be further
suppressed as $\Lambda_{inst}^4 = M^2 \mu^2 \exp(-S_{inst})$, where
$\mu = \sqrt{m_{3/2} M_P}$ is the SUSY breaking scale. In order to
have the successful PQ mechanism, it is usually assumed that the
action $S_{inst}$ is very large (e.g. $S \simeq 200$), which
suppresses the explicit breakings small enough. For our purpose, $S$
should not be that large, since we need the large explicit breaking
terms.  Since the size of the breaking $\Lambda_{inst}^4$ is very
sensitive to $S_{inst}$, it is important to estimate the value of
$S_{inst}$ very precisely. We here
simply treat $\Lambda_{inst}$ (therefore $m_a$) as a free parameter.

First let us consider a case that the axion mass is heavier than the
cosmic expansion rate during inflation, i.e., $m_a > H_{inf}$. Then
the axion settles down to the potential minimum during
inflation. Since the anthropic argument requires the minimum to
coincide with the CP conserving one, the axion remains to stay there
after inflation, and the cosmological abundance of the axion is
negligible.  Therefore, in this case, the axion does not play any
important role in cosmology.

Next we take up the other case that the axion mass is lighter than the
cosmic expansion rate during inflation. Then the position of the axion
during inflation is expected to be away from the CP conserving minimum
by ${\cal O}(1)$.  After inflation, the axion starts to oscillate when
the Hubble parameter becomes comparable to the axion mass. What is
different from the ordinary PQ mechanism is that the oscillations can
start in much earlier phase of the universe, and more importantly,
that the axion is unstable and decays into the SM particles.

The partial decay rate of the axion into a pair of
the gluons through (\ref{eq:axion-gluon}) is given by
\beq
\Gamma(a \rightarrow 2 g) \;\simeq\; \frac{\alpha_s^2}{64 \pi^3} \frac{m_a^3}{f_a^2}.
\eeq
Assuming that the possible decay processes into the other sectors are
kinematically forbidden, 
the decay temperature of the axion, $T_a$,  is
\beq
T_a \;\simeq\; 8 \times 10^7{\rm\,GeV} \lrfp{g_*}{200}{-\frac{1}{4}} 
\lrf{\alpha_s}{0.05} \lrfp{m_a}{10^{12}\,{\rm GeV}}{\frac{3}{2}}
\lrfp{f_a}{10^{16}{\rm\,GeV}}{-1},
\eeq
where $g_*$ counts the relativistic degrees of freedom at the decay.
Since the initial amplitude of the axion is as large as $f_a = {\cal
O}(10^{16})\,$GeV, the axion abundance tends to be quite
large. Therefore the axion must decay before the big bang
nucleosynthesis (BBN) starts. Requiring $T_a \gtrsim
10{\rm\,MeV}$~\cite{Kawasaki:1999na}, the axion mass is bounded below:
\beq
m_a \;\gtrsim\; 2 \times 10^5{\rm \,GeV} \,
\lrfp{g_*}{200}{\frac{1}{6}} 
\lrfp{\alpha_s}{0.05}{-\frac{2}{3}}
\lrfp{f_a}{10^{16}{\rm\,GeV}}{\frac{2}{3}}.
\eeq
The explicit breaking of the shift symmetry should be large enough
that this condition is met when $m_a \lesssim H_{inf}$. Note that 
the lower limit is not applied in the case of $m_a \gtrsim H_{inf}$.

The cosmological abundance of the axion is estimated to be
\beq
\frac{\rho_a}{s} \;\simeq\; \frac{1}{8} \,T_{inf} \lrfp{f_a}{M_P}{2},
\eeq
where $\rho_a$ is the energy density of the axion, $s$ the entropy
density, and $T_{inf}$ the inflaton decay temperature. We have here
assumed that the initial displacement of the axion from the minimum
is equal to $1$, and that the
axion does not dominate the energy density of the universe.  This is
the case if $T_{inf} \lesssim 6 T_a (M_P/f_a)^2$. On the other hand,
if $T_{inf} \gtrsim 6 T_a (M_P/f_a)^2$, the axion dominates the
universe, and the (last) reheating of the universe is provided by the
decay of the axion.

The latter possibility is particularly interesting. The reheating
temperature of the universe is completely determined by the parameters
of the axion sector, i.e., $\Lambda_{inst}$ and $f_a$, which are in
principle calculable once the axion model is fixed in the string
theory. Furthermore, since the axion is light during inflation, it
acquires quantum fluctuations, which turn into the adiabatic density
perturbations after the decay~\footnote{If the axion does not dominate
the energy density of the universe, it may be able to generate large
non-Gaussianity either by the curvaton mechanism~\cite{Lyth:2006gd} or
by the ungaussiton mechanism~\cite{Linde:1996gt,Suyama:2008nt}.
}. That is, the axion can be a curvaton~\cite{curvaton}, if the
inflation scale is $H_{inf} \sim 10^{-5} (2\,\pi f_a) \sim
10^{12}$\,GeV.  Such a paradigm may help us to construct an inflation
model in the stringy set-up, because the density perturbations do not
have to be generated by the inflaton, and because the reheating is
naturally induced by the axion that has a couplings to the SM sector.

We make several remarks on the other cosmological implications.  Note
that, since there is no need to dilute the axion, the attractive
cosmological scenarios such as the leptogenesis~\cite{Fukugita:1986hr}
are feasible for large enough $m_a$. Even if the axion mass is relatively small and
the reheating temperature due to the axion decay becomes rather low,
we do not need to introduce another sector in order to dilute the axion.
In our scenario, the axion does not contribute to the
DM abundance, which suggests that other candidates such as WIMP and
the gravitino should account for the DM. The failure of the anthropic argument
to account for both the cosmological constant and the DM abundance simultaneously
may hint that the DM abundance is determined by the physics, not by the anthropic
reasoning. 
Also, since the saxion mass is also large, its cosmological problem can be solved
in a similar fashion.  
In the discussion above, we have simply assumed that the axion mainly decays into a pair
of the gluons. There might be other decay processes at
tree-level~\cite{MGP, Endo:2006tf, Endo:2006qk,Endo:2006xg} as well as
one-loop level~\cite{Endo:2007ih}. In particular, the gravitino might
be non-thermally produced by the axion decay, which may help us
further constrain the axion models.

%%%%%%%%%%%%%%%%%%%%
\section{Conclusions and discussion}
\label{sec:4}
%%%%%%%%%%%%%%%%%%%%
The existence of the axion-like fields is quite common in the string
theory. They generically receive explicit breakings of the
shift symmetries due to the world-sheet instantons, brane instantons,
gauge instantons from other factors of the gauge group, and
gravitational instantons.  It was indeed an issue how to suppress such
explicit breakings in order to have the successful PQ
mechanism~\cite{Svrcek:2006yi}.  This generically set a restriction on
the theory. We have offered a possibility to solve the strong CP
problem in the presence of large explicit breaking terms, which
therefore liberate the theory from such restriction.  An essential ingredient 
is an assumption that the probability distribution of the vacuum energy
excluding the contribution from the axion sector has a pressure towards higher values.
Then, among those vacua satisfying the anthropic bound on the cosmological constant,
the CP conserving minimum is statistically favored since it minimizes the contribution
from the axion sector. Note that such a probability distribution can be consistent with
a flat prior usually assumed when one applies the anthropic principle to the
cosmological constant problem, since the energy scales of interest are different.

At present we do not know the origin of the steep probability distribution. 
It is interesting to note,
however, that there are some proposals that the distribution might differ from the
flat distribution~\cite{ArkaniHamed:2005yv,SchwartzPerlov:2006hi,
SchwartzPerlov:2006hz,Olum:2007yk}.   The source of
the hierarchy may be the statistical property of the scanning of $ \rho_{ L} $
and/or some dynamics such as the bubble nucleation. The measure of the
distribution of vacua, taking account of the cosmic expansion during eternal inflation,
may also help us to understand the origin of such hierarchy.

It would be encouraging if we can find other
examples in which such steep vacuum distribution plays 
an important role to determine  physical environmental parameters.
In particular, if the the probability distribution is steep
over a scale of the weak scale or larger, it will favor a heavier Higgs mass,
since it results in the deeper potential well for the fixed Higgs VEV. Thus,
it will be quite interesting and suggestive,
if the little hierarchy problem associated with the Higgs mass can be 
interpreted as the result of such vacuum distribution. Of course we need
to properly take account of the anthropic window on the electroweak breaking
scale, in order to claim that the steep vacuum distribution favors 
large one-loop corrections to the Higgs mass.

Throughout this letter we have not specified the source for the
explicit breaking. If it is large enough, the axion will settle down
at the CP conserving minimum during inflation. Thus the cosmological
abundance of the axion is negligible in this case. It is also possible
that the axion dominates the energy density of the universe after
inflation and reheats the universe by the decay, if the explicit
breaking is relatively small during inflation.

How large the explicit breaking can be in the realistic string theory
and its implication on the inflationary scale are very interesting
issues, and we leave them for future work.  
 Whether a steep probability distribution is indeed feasible or not, as
well as how much hierarchy can be realized and from what it is
originated, are open questions.  Hopefully,
future development in the string theory and the associated areas 
may enable us to answer all or some of these questions.

%%%%%%%%%%%%%%%%%%%%%%%%%%%%%%%%
%\section{Discussion and Conclusions} 
%\label{sec:4}
%%%%%%%%%%%%%%%%%%%%%%%%%%%%%%%%

\bigskip
\bigskip

\noindent {\bf Acknowledgments:} F.T. would like to thank 
S.~Hellerman, M.~Kawasaki, K.~Nakayama and R. Bousso for discussion. 
   This work was supported by World Premier International
   Research Center InitiativeiWPI Initiative), MEXT, Japan.

\end{document}